\documentclass[%
 reprint,
 amsmath,amssymb,
 aps,
prl
]{revtex4-1}

\usepackage{lmodern}
\usepackage{graphicx}
\usepackage{dcolumn}
\usepackage{bm}
\usepackage{hyperref}
\hypersetup{linktocpage,colorlinks,citecolor={blue},pdfdisplaydoctitle=true,pdfpagemode=UseOutlines,bookmarksnumbered=true}
\usepackage{mathrsfs,dsfont}
\usepackage{color}

\makeatletter
\renewcommand{\@biblabel}[1]{[#1]\hfill}
\makeatother

\usepackage[normalem]{ulem}

\newcommand{\upd}{\mathrm{d}}
\newcommand{\tr}{\mathrm{tr}}
\newcommand{\ie}[0]{\textit{i.e.} }
\newcommand{\eg}[0]{\textit{e.g.} }

\newcommand{\rholin}[0]{\widetilde{\rho}}
\newcommand{\rhoj}[0]{\rho_{\mathbf{j}}}

\newcommand{\rhoavj}[0]{\bar{\rho}_\mathbf{j}}

\begin{document}
\title{Exact signal correlators in continuous quantum measurement}
\author{Antoine Tilloy}
\email{antoine.tilloy@mpq.mpg.de}
\affiliation{Max-Planck-Institut f\"ur Quantenoptik, Hans-Kopfermann-Stra{\ss}e 1, 85748 Garching, Germany}
\date{\today}

\begin{abstract}
This article provides an exact formula for the signal $n$-point correlation functions of detectors continuously measuring an arbitrary quantum system, in the presence of detection imperfections.  The derivation uses only continuous stochastic calculus techniques, but the final result is easily understood from a discrete picture of repeated interactions with qubits or from a parallel with continuous matrix product states. This result provides a crude yet efficient a way to estimate system parameters directly from experimental data, without requiring non-linear state reconstruction.
\end{abstract}

\maketitle

\paragraph{Introduction --}
Quantum systems can now be probed non-destructively with the help of continuous measurements. Intuitively, the latter can be understood as coming from a proper limit of infinitely frequent and infinitely weak discrete measurements \cite{attal2006,attal2010}, where the scaling is chosen in such a way that interactions extract information from the state without Zeno-freezing its evolution in the limit. The corresponding theory was developed in the 80s \cite{mensky1979,barchielli1982,barchielli1986,caves1986,diosi1988,barchielli1991}, mostly with foundational motivations. However, it was later understood that continuous measurements could be obtained in standard experimental setups \cite{wiseman1993,wiseman2009}, making the construction practically relevant. Nowadays, continuous measurements are almost routinely used by experimentalists and the corresponding continuous quantum trajectories can be reconstructed, notably in superconducting qubits \cite{vijay2012, katz2006,campagne2016}. The interest is manifest: by acting on a quantum system conditionally on a continuous measurement readout (or signal), it is possible to control its state in real time. This can be used to enable or optimize tasks such state preparation \cite{vanhandel2005}, purification \cite{jacobs2003,combes2006}, rapid measurement \cite{combes2008,combes2011,tilloy2016}, or quantum error correction \cite{ahn2002,atalaya2017_0}.

The theory typically outputs an expression for the system state $\rho_t$ as a non-trivial function of the history of the measured signal $I_t$. 
In practice, this is not very convenient to test the theory or fit unknown parameters. Indeed, the continuously evolving state itself is not a quantity one can directly measure to compare theory and experiment. Although the system parameters can still be estimated using maximum likelihood methods \cite{chase2009,gambetta2001}, this requires solving a non-linear filtering equation with a discretized signal \cite{six2015}.
However, there is a much simpler (albeit mathematically suboptimal) option which consists in looking only at basic functions of the signal, like $n$-point correlation functions \cite{jordan2005,atalaya2017_0,atalaya2017_1}.
Such quantities are trivial to measure in practice: they are direct multiplications of (delayed) detector outputs and can thus be obtained analogically. The non-trivial step becomes to compute the corresponding theoretical prediction. 

Fortunately, an exact expression, taking measurement imperfections into account, can be found using standard stochastic calculus techniques. It generalizes old results of continuous measurement theory  \cite{zoller1997,barchielli2006,barchielli2009} as well as recent rediscoveries \cite{jordan2005,diosi2016,atalaya2017_1,atalaya2017_2} valid in restricted cases. The main objective of this article is to provide a self-contained derivation and an interpretation of this formula, as well as an example illustrating its convenience for parameter estimation.

\paragraph{Standard formalism --}
We consider a quantum system subjected to a continuous measurement by $n$ independent detectors. A general \footnote{The SME can actually take an even more general form (see \eg \cite{wiseman2001,barchielli1993,barchielli1995}). Our derivation is straightforward to extend to such settings.} stochastic master equation (SME) in It\^o form, describing the real-time evolution of the system density matrix $\rho$ under this continuous probing, can be shown to read \cite{wiseman2009,jacobs2006}:
\begin{equation}\label{eq:sme}
    \upd \rho = \mathcal{L}(\rho) \upd t + \sum_{k=1}^n \mathcal{D}[c_k] (\rho)\,\upd t + \sqrt{\eta_k} \, \mathcal{H}[c_k](\rho) \,\upd W_k,
\end{equation}
where the $W_k$ are independent Wiener processes (or Brownian motions), $\mathcal{L}$ is an arbitrary Lindbladian encoding the evolution of the system in the absence of measurement, $0<\eta_k \leq 1$ are the detector efficiencies, and
\begin{align}\label{eq:notation}
\mathcal{D}[c](\rho)&=c\rho c^\dagger-\frac{1}{2}\left\{c^\dagger c,\rho\right\},\\
\mathcal{H}[c](\rho)&=c\rho + \rho c^\dagger - \tr\left[(c+c^\dagger)\rho\right]\rho,
\end{align}
where the $c_k$ are arbitrary operators characterizing each detector.
The corresponding measurement signals $I_k(t) = \frac{\upd r_k(t)}{ \upd t}$ verify \footnote{We have taken the convention of \cite{jacobs2006} which means there is a factor of $2$ difference with the definition of the signal in \cite{atalaya2017_2}.}:
\begin{equation}\label{eq:signal}
\upd r_k = \frac{1}{2}\tr[(c_k+c^\dagger_k)\rho] \, \upd t +\frac{1}{2 \sqrt{\eta_k}} \upd W_k.
\end{equation}
This latter equation can be used to express $\upd W_k$ as a function of $\upd r_k$ and thus to reconstruct $\rho$ as a function of the signal using \eqref{eq:sme}.

We are interested in correlation functions of the signal:
\begin{equation}
   K_{\ell_1\ell_2\cdots\ell_N}(t_1,t_2,\cdots,t_N):= \mathds{E}\left[I_{\ell_1} (t_1) I_{\ell_2} (t_2) \cdots I_{\ell_N}(t_N)\right],
\end{equation}
where $\mathds{E}$ denotes the statistical average over the Brownian randomness.
Our objective is to compute these correlators analytically using only equations \eqref{eq:sme} and \eqref{eq:signal}.

For that matter, it is mathematically convenient go to a linear SME instead of a non-linear one. Introducing $\rholin$ such that: 
\begin{equation}\label{eq:linearSME}
    \upd \rholin = \mathcal{L}(\rholin) \upd t + \sum_{k=1}^n \mathcal{D}[c_k](\rholin) \upd t + 2 \,\eta_k\left[c_k \rholin + \rholin c^\dagger_k\right]\upd r_k
\end{equation}
and $\rholin(0)=\rho(0)$ ,
it is easy to show, using It\^o's lemma, that $\rho(t)=\rholin(t) \times \tr[\rholin(t)]^{-1}$. We further define a new probability measure $ \widetilde{\mathds{P}}$ from the physical probability measure $\mathds{P}$ by the Radon-Nikodym derivative:
\begin{equation}
    \frac{\upd \widetilde{\mathds{P}}}{\upd \mathds{P}} = \tr\{\rholin(T)\}^{-1},
\end{equation}
for a fixed time $T$ arbitrarily far in the future (which we here take to be larger than all the time arguments in the correlators).
In practice, it amounts to introduce a new expectation value:
\begin{equation}\label{eq:transformed}
    \widetilde{\mathds{E}}\Big[\;\cdot \; \Big] = \mathds{E}\Big[\;\cdot \; \tr\{\rholin(T)\}^{-1}\Big].
\end{equation}
Under this new measure, one can show (using Girsanov's theorem \cite{oksendal2003}) that the $2\sqrt{\eta_k} r_k$ are independent Wiener processes for time arguments $\leq T$ \footnote{This is sometimes done the other way around (see e.g. \cite{jacobs2006,barchielli2009}). Starting from a linear SME with Gaussian signal, one derives the non-linear SME \eqref{eq:sme} and drifted signal \eqref{eq:signal} normalizing the solution and changing the probability measure.}. In the end, this means the correlators can be computed by writing:
\begin{equation}\label{eq:correlatorsimplified}
   K_{\ell_1\cdots\ell_N}(t_1,\cdots,t_N)= \widetilde{\mathds{E}}\left[I_{\ell_1} (t_1)  \cdots I_{\ell_N}(t_N)\times \tr\{\rholin(T)\}\right],
\end{equation}
where $\rholin$ now obeys the \emph{linear} equation \eqref{eq:linearSME} and the $I_k$'s are just independent white noises.

\paragraph{Derivation --} 
The main idea is to introduce the generating functional $\mathcal{Z}_\mathbf{j}$ of the correlation functions:
\begin{equation}
    \mathcal{Z}_\mathbf{j}:=\mathds{E}\left[\exp\left(\sum_{k=1}^n \int_0^T j_k(u) \,\upd r_k(u) \right)\right],
\end{equation}
where $\mathbf{j}=(j_k)_{k=1}^n$ is a ``source'' function and as before, $T$ is kept finite to make the expression well defined but should typically be sent to infinity in the final results (or at least taken to be larger than all the other time arguments).
Assuming for the time being that $t_1<t_2<\cdots < t_N$, we indeed have:
\begin{equation}\label{eq:functionaldiff}
K_{\ell_1\cdots\ell_N}(t_1,\cdots,t_N) 
     = \frac{\delta}{\delta j_{\ell_1}(t_1)}\cdots \frac{\delta}{\delta j_{\ell_N}(t_N)} \mathcal{Z}_\mathbf{j}\Big|_{\mathbf{j}=0},
\end{equation}
something noted also in \cite{barchielli2009}.
Introducing $\rhoj(t) = \exp\left(\sum_{k=0}^n \int_0^t j_k(u) \,\upd r_k(u) \right)\rholin(t)$, we can write: $\mathcal{Z}_\mathbf{j}=\widetilde{\mathds{E}}\left[\tr \left\{\rhoj(T)\right\}\right]$. Using It\^o's lemma, we now compute the differential of $\rhoj$:
\begin{equation}\label{eq:smefunctional}
\begin{split}
    \upd \rhoj= &\mathcal{L}(\rhoj) \upd t+ \mathcal{D}[c_k](\rhoj) \upd t+2 \eta_k\left[c_k \rhoj + \rhoj c^\dagger_k\right]\! \upd r_k\\
    +& \frac{j_k^2}{8\eta_k} \rhoj\,\upd t +\frac{j_k}{2}  \left[c_k \rhoj + \rhoj c^\dagger_k\right]\upd t + j_k \rhoj \, \upd r_k,
\end{split}
\end{equation}
with implicit summation on $k$. Because the $r_k$ are proportional to Wiener processes and because we are in the It\^o representation, the average of all stochastic integrals taken against $\upd r_k$ is zero. Hence, writing $\rhoavj=\widetilde{\mathds{E}}[\rhoj]$, and taking the expectation value of \eqref{eq:smefunctional}, we get:
\begin{equation}\label{eq:generatingdiff}
    \partial_t \rhoavj= \left(\mathcal{L} + \sum_{k=1}^n\mathcal{D}[c_k] + \frac{j_k}{2} c_k^+ + \frac{j_k^2}{8\eta_k}\right) \cdot \rhoavj
\end{equation}
with the notation  $c^+ \cdot \rho= c \rho + \rho c^\dagger$. The differential equation \eqref{eq:generatingdiff} can be formally integrated to give:
\begin{equation}\label{eq:rhoj}
    \rhoavj(T)= \mathcal{T}\!\exp\!\left(\int_0^T\!\!\!\upd t\, \mathcal{L} + \sum_{k=1}^n\mathcal{D}[c_k] + \frac{j_k}{2} c_k^+ + \frac{j_k^2}{8\eta_k}  \right)  \rho(0)
\end{equation}
where $\mathcal{T}$ is the time ordering operator. Taking the trace and using \eqref{eq:functionaldiff} we get:
\begin{equation}\label{eq:correlatorregular}
    \begin{split}
        &K_{\ell_1\ell_2\cdots\ell_N}(t_1,t_2,\cdots,t_N) = \frac{1}{2^N}   \\
        &\times \tr \left[c^+_{\ell_N}  \Phi_{t_N - t_{N-1}}   c^+_{\ell_{N-1}}
\cdots \, \Phi_{t_2 - t_{1}} c_{\ell_1}^+ \Phi_{t_1} \cdot \rho(0)\right],
    \end{split}
\end{equation}
where $\Phi_t=\exp\left(t\mathscr{L}\right)=\exp\left\{t\left( \mathcal{L} + \sum_{k=1}^n\mathcal{D}[c_k] \right)\right\}$ is just the solution of a Lindblad evolution if measurements are averaged over (the extension to time dependent Lindblad is straightforward). 

Formula \eqref{eq:correlatorregular} has a simple interpretation in terms of discrete positive-operator valued measure (POVM). Let us introduce $n$ POVMs $F_k$ with binary outcomes $R_k=\pm 1$ such that:
\begin{equation}
    F_k:\rho\mapsto \frac{M_k(\pm 1) \rho M_k^\dagger(\pm 1)}{\tr [M_k(\pm 1) \rho M_k^\dagger(\pm 1)]}, 
\end{equation}
with probability $\mathds{P}[R_k=\pm 1] =\tr [M_k(\pm 1) \rho M_k^\dagger(\pm 1)]$ and
\begin{align}
    M_{k}(R_k=+1) &=\frac{1}{\sqrt{2}} \left(\mathds{1} + \delta c_{k} - \delta^2  \frac{c_k^\dagger c_k}{2} +o(\delta^2)\right)\\ 
    M_{k}(R_k=-1)&=\frac{1}{\sqrt{2}} \left(\mathds{1} - \delta  c_{k} - \delta^2  \frac{c_k^\dagger c_k}{2} + o(\delta^2)\right)\!\!.
\end{align}
We now consider a continuous evolution given by $\Phi_t$ interrupted by the application the POVM $F_{\ell_k}$ at discrete times $t_k$, and write $R_{\ell_k}(t_k)$ the associated outcome. The correlation function $K$ we have derived is simply the correlation function of the discrete binary results obtained in this scheme. Indeed, after elementary algebra one gets:
\begin{align}
\frac{\mathds{E}[R_{\ell_1}(t_1) \cdots R_{\ell_N} (t_n)]}{\delta^N}\underset{\delta\rightarrow 0}{\sim}&
    \tr \left[c^+_{\ell_N}  \Phi_{t_N - t_{N-1}}  \!\!\! \cdots c_{\ell_1}^+ \Phi_{t_1}  \rho(0)\right] \nonumber\\
    =&2^{N}K_{\ell_1, \cdots,\ell_N}(t_1,\cdots,t_N) 
\end{align}
This reformulation (observed already in \cite{atalaya2017_2} in a special case) is natural having in mind that continuous quantum measurements can be obtained as a limit of repeated interactions with qubits \cite{tilloy2014,gross2018}. It incidentally shows how formula \eqref{eq:correlatorregular} could have been derived from the discrete.

\paragraph{Equal point contributions --}
We recall that formula \eqref{eq:correlatorregular} is valid only for $t_1<\cdots<t_N$. Indeed functional derivatives of $\mathcal{Z}_\mathbf{j}$ taken with respect to the same $j_k$ and at the same time  would have yielded Dirac distributions, corresponding to the singular two-point functions of white noises. It is important not to miss these contributions: in practice the signals are obtained from amplifiers with a finite bandwidth and are thus effectively filtered. If one computes correlation functions for time differences smaller than the inverse cutoff frequency of the amplifiers, these contributions will be significant.
In practice the signals are well defined once smoothed with a test function $f$ (which, experimentally, would correspond to the transfer function of the amplifier taken at a fixed time):
\begin{equation}
    I_k(f) := \int \!\! f(t) \, \upd r_k(t).
\end{equation}
Consequently, it is actually more appropriate to consider that the correlation functions also act on test functions:
\begin{equation}
K_{\ell_1,\cdots,\ell_N}(f_1,\cdots,f_N) := \mathds{E}\left[I_{\ell_1}(f_1)\cdots I_{\ell_N}(f_N)\right].
\end{equation}
To compute such a correlation functional applied on test functions, we introduce a modified generating function $\mathcal{Z}_{\varepsilon,\mathbf{f}}$, which is just equal to $\mathcal{Z}_\mathbf{j}$ for $j_k=\sum_{i=1}^N\varepsilon_i f_i \delta_{\ell_i,k}$:
\begin{equation}
    \mathcal{Z}_{\varepsilon,\mathbf{f}}=\mathds{E}\left[ \exp\left(\sum_{i=1}^N\int_0^T \! \varepsilon_i \, f_i(u) \,  \upd r_{\ell_i}(u)\right)\right].
\end{equation}
It is indeed such that:
\begin{equation}
    K_{\ell_1,\cdots,\ell_N}(f_1,\cdots,f_N) = \partial_{\varepsilon_{1}} \cdots \partial_{\varepsilon_{N}}\mathcal{Z}_{\varepsilon,\mathbf{f}} \Big|_{\varepsilon=0}.
\end{equation}
Using \eqref{eq:rhoj}, we have:
\begin{equation}\label{eq:alternategenerating}
\begin{split}
&\mathcal{Z}_{\varepsilon,\mathbf{f}}=\underset{[\spadesuit]}{\underbrace{\exp\left(\frac{1}{8 \, \eta_{\ell_i}}\int \varepsilon_i\varepsilon_{i'} \, \delta_{\ell_i\ell_{i'}}\, f_i f_{i'} \right) }} \times\\
&\underset{[\clubsuit]}{\underbrace{\tr \bigg[\mathcal{T}\exp\bigg(\int_0^T\!\!\!\mathcal{L} + \sum_{k=1}^n\mathcal{D}[c_k] + \frac{\varepsilon_i f_i}{2} c_{\ell_i}^+  \bigg)\cdot \rho(0) \bigg]}},
\end{split}
\end{equation}
with implicit summation on $i$ and $i'$. To compute correlation functions, we just need to apply the derivatives with respect to the $\varepsilon$'s on the product $\spadesuit\times\clubsuit$. Let us consider the second term $\clubsuit$ first. The derivatives with respect to the $m$ first $\varepsilon$'s give for example:
\begin{align}
\partial_{\varepsilon_1}\cdots \partial_{\varepsilon_m}  \clubsuit \big|_{\varepsilon=0}&=\tr\Bigg(\mathcal{T}\bigg\{ \prod_{i=1}^m\int_0^T\!\!  \frac{f_i(t_i) }{2}c_i^+ \,\upd t_i \nonumber\\
\times \exp& \left[\int_0^T \! \mathcal{L} + \sum_{k=1}^N \mathcal{D}[c_{\ell_k}]\right] \bigg\}\cdot \rho(0)\Bigg).
\end{align}
Expanding the product, pulling the integrals out of the time ordering and using equation \eqref{eq:correlatorregular} then yields:
\begin{align}\label{eq:deriv1}
&\partial_{\varepsilon_1}\cdots \partial_{\varepsilon_m}  \clubsuit \Big|_{\varepsilon=0}=\sum_{\sigma \in \mathscr{S}^m} \int_{t_1\leq \cdots \leq t_m} \!\!\!\!\upd t_1 \cdots \upd t_m \nonumber\\ 
&\times f_{{\sigma(1)}}(t_1) \cdots f_{{\sigma(m)}} (t_m)
\, K_{\ell_{\sigma(m)}\cdots\ell_{\sigma(m)}}(t_1,\cdots,t_m)\\
&:= K^\circ_{\ell_1 \cdots \ell_m}(f_{1},\cdots, f_{m}),
\end{align}
where $\mathscr{S}^m$ is the set of permutations of $\{1,2,\dots,m\}$ and $K(t_1,\cdots,t_m)$ is given by \eqref{eq:correlatorregular}.
The right hand side of \eqref{eq:deriv1} corresponds to the natural definition of the correlator  without singular contributions \eqref{eq:correlatorregular}, but for smoothed signals. We now compute the result of $2m$ derivatives acting on the first term $\spadesuit$ of the product in \eqref{eq:alternategenerating} :
 \begin{equation}\label{eq:deriv2}
 \begin{split}
\partial_{\varepsilon_1}\cdots \partial_{\varepsilon_{2m}} \spadesuit\Big|_{\varepsilon=0}\!\!= \!\!  \sum_{P \in \mathscr{P}^{2m}} \prod_{(p,p')\in P} \frac{\delta_{\ell_p\ell_{p'}}}{4 \eta_{\ell_p}} \!\!\int_0^T \!\!\!f_p f_{p'}
 \end{split}
 \end{equation}
where $(p,p')$ is a pair of indices, $P$ is a pairing, \ie a set of $m$ pairs of $2m$ indices, and $\mathscr{P}^{2m}$ is the set of all the $m!!$ possible pairings. This amounts to sum over all the possible products of pairwise ``contractions" of test functions associated to the same detector. This is simply Wick's theorem in disguise for the equal point contributions.

To get the final result, we just need to distribute the derivatives in \eqref{eq:alternategenerating}: 
\begin{equation}
\setlength{\jot}{-8pt}
\begin{split}
    \partial_{\varepsilon_1}\!\!\cdots \partial_{\varepsilon_N} (\spadesuit \! \times \! \clubsuit)= \!\!\!\!\!\!\!\sum_{\underset{\in \{0,1\}^N}{(s_1,\cdots,s_N)}}& \partial_{\varepsilon_{\alpha_s(1)}} \!\!\cdots \partial_{\varepsilon_{\alpha_s(|s|)}} \spadesuit \\ 
   \underset{\underset{~}{~}}{~} &\times  \partial_{\varepsilon_{\beta_s(1)}} \!\!\cdots \partial_{\varepsilon_{\beta_s(N-|s|)}}\clubsuit,
    \end{split}
\end{equation}
where $|s|=s_1 + \cdots s_N$, $\alpha_s$ outputs the $s$ indices $i$ such that $s_i=1$, and $\beta_s$ outputs the $N-|s|$ indices such that $s_i=0$. Finally, using \eqref{eq:deriv1} and \eqref{eq:deriv2} to compute each side we obtain the index heavy (albeit fully explicit) formula:
\begin{widetext}
\begin{equation}\label{eq:fullcorrelator}
\begin{split}
K_{\ell_1,\cdots,\ell_N}(f_1,&\cdots,f_N) = K^\circ_{\ell_1 \cdots \ell_N}(f_1,\cdots, f_N)\\
+&\sum_{\underset{\in \{0,1\}^N}{(s_1,\cdots,s_N)}} \sum_{P \in \mathscr{P}^{|s|}} \prod_{(p,p')\in P}\left[\frac{\delta_{\ell_{\alpha_s(p)}\ell_{\alpha_s(p')}}}{4 \eta_{\ell_{\alpha_s(p)}}} \!\!\int_0^T \!\!\!f_{\alpha_s(p)} f_{\alpha_s(p')}\right]K^\circ_{\ell_{\beta_s(1)} \cdots \ell_{\beta_s(N-|s|)}}(f_{{\beta_s(1)}},\cdots, f_{{\beta_s(N-|s|)}}),
\end{split}
\end{equation}
where we recall that:
\begin{equation}
    K^\circ_{\ell_1 \cdots \ell_k}(f_{1},\cdots, f_{k})=\!\!\sum_{\sigma \in \mathscr{S}^k} \int_{t_1\leq \cdots \leq t_k} \!\!\!\! \!\!\!\!\!\!\!\!\!\!\!\!\upd t_1 \cdots \upd t_k
\;f_{{\sigma(1)}}(t_1) \cdots f_{{\sigma(k)}} (t_k) \; \tr \left[c^+_{\ell_k}  \Phi_{t_k - t_{k-1}}   c^+_{\ell_{k-1}}
\cdots \, \Phi_{t_2 - t_{1}} c_{\ell_1}^+ \Phi_{t_1} \cdot \rho(0)\right].
\end{equation}
\end{widetext}
In practice, most terms in the sum \eqref{eq:fullcorrelator} will be zero as only products of contractions from the same detectors contribute. It will thus typically be more convenient to use \eqref{eq:alternategenerating}, \eqref{eq:deriv1}, and \eqref{eq:deriv2} to compute directly the non-zero terms. In the limit of test functions with non-overlapping supports, all the terms containing contractions vanish and $K_{\ell_1,\cdots,\ell_N}=K^\circ_{\ell_1,\cdots,\ell_N}$. The same identity naturally applies if all the signals in the correlator come from different detectors. On the other hand, correlators with equal point contributions are particularly useful as they depend explicitly on the efficiencies $\eta_k$, and thus allow to estimate them.

\paragraph{Example --} One of the simplest example one can consider is the simultaneous measurement of two non-commuting qubit operators, a situation inspired from \cite{ficheux2017}. We take a qubit of density matrix $\rho$, without proper dynamics ($\mathcal{L}=0$), continuously measured with  $c_1=\sqrt{\gamma_x} \sigma_x$ (with associated signal $I_1:=I_x$) and $c_2=\sqrt{\gamma_{-}} \sigma_- = \sqrt{\gamma_{-}}(\sigma_x-i\sigma_y)/2$ (with associated signal $I_2:=I_-$)
where $\sigma_x$, $\sigma_y$, and $\sigma_z$ are the Pauli matrices. 

This example is non-trivial because the measured operators do not commute and the evolution is not even unital. But it is also convenient because the map $\Phi_t$ --corresponding to the evolution averaged over the measurement outcomes-- is easily diagonalized by writing $\rho$ in the basis of the Pauli matrices. For simplicity, we consider a cross correlation function $K_{x,-}(t_1,t_2)=\mathds{E}[I_x(t_1) I_-(t_2)]$ that has no equal time contribution and that we thus write as a function for convenience. Using the general formula \eqref{eq:fullcorrelator} and elementary algebra yields:
\begin{equation}\label{eq:exactexample}
    \begin{split}
        &K_{x,-}(t_1,t_2) =\frac{\sqrt{\gamma_-\gamma_x}}{2} \, \mathrm{e}^{-\gamma_-|t_2-t_1|/2}\times\bigg\{\theta_{t_2 t_1}+\\
        &\theta_{t_1 t_2}\!\left[ \frac{2\gamma_x}{\gamma_-\!\!+2\gamma_x}  -\left(z_0-\frac{\gamma_-}{\gamma_-\!\!+2\gamma_x}\right)\mathrm{e}^{-(\gamma_- \!+ 2\gamma_x) t_1}\right]\bigg\},
    \end{split}
\end{equation}
where $z_0=\tr[\sigma_z \rho(0)]$ and $\theta_{t_1 t_2}$ is the Heaviside function, equal to $1$ if $t_1>t_2$ and $0$ otherwise.

This formula illustrates several generic features of signal correlation functions. The function $K_{x,-}$ is asymmetric in the exchange $t_1 \leftrightarrow t_2$. It depends on the initial state although it forgets it for large times. It decreases exponentially quickly in $|t_1-t_2|$. It is discontinuous in $t_1=t_2$, reminding us that correlation functions only make sense as distributions: the value in $t_1=t_2$ depends on the regularization procedure, i.e. on the way the signals are filtered. Finally, the interest of our general formula for parameter estimation is made obvious: $\gamma_-$ and $\gamma_x$ can unequivocally be read from $K_{x,-}$.

\begin{figure}
    \centering
    \includegraphics[width=0.98\columnwidth]{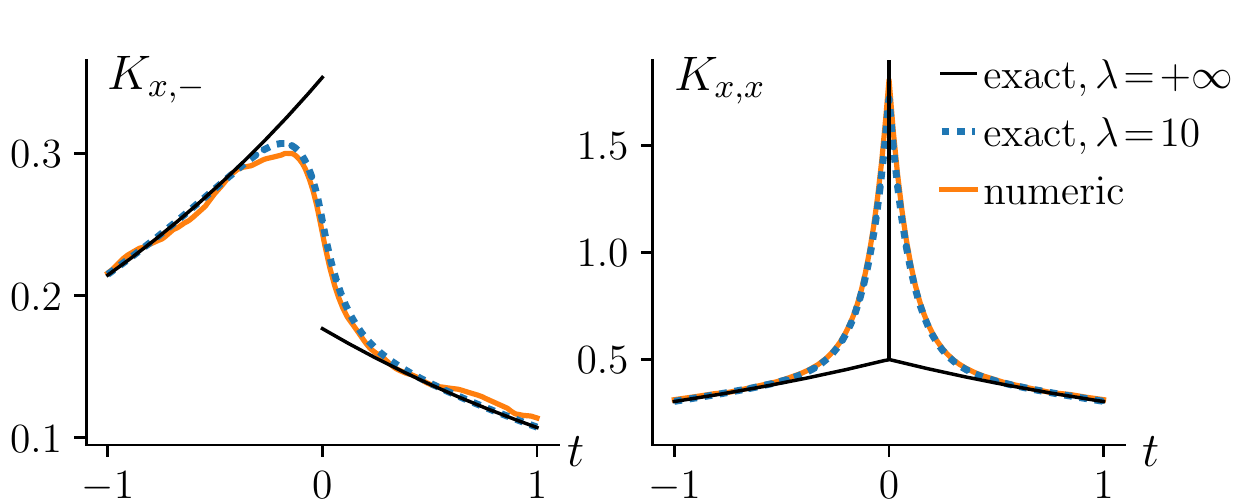}
    \caption{Correlations functions $K_{x,-}(f^t,f^0)$ and $K_{x,x}(f^t,f^0)$. The time unit is $[T]=\gamma_-^{-1} \equiv 1.0 $, and $\gamma_x=0.5$, $\eta=1.0$. Lower efficiencies do not impact $K_{x,-}$, but enhance the central peak of $K_{x,x}$. Numerics are obtained for one trajectory of length $10^4 [T]$ (voluntarily short), computed from \eqref{eq:sme} and \eqref{eq:signal} with a naive Euler scheme and $\delta t = 10^{-3}$, filtered with $\lambda=10$. }
    \label{fig:correlations}
\end{figure}

We may further illustrate how the imperfections of the amplification chain can be taken into account. A smoothing of the signal by a typical first order filter corresponds to a test function $f^t:u\mapsto \theta_{t,u} \lambda \exp[-\lambda (t-u)]$ where $\lambda$ is the bandwidth. Using \eqref{eq:exactexample} we can easily compute $K_{x,-}(f^t,f^0)$ analytically in the stationary state. The resulting correlation function is shown in Fig. \ref{fig:correlations} alongside $K_{x,x}$ which can be obtained similarly. The smoothing removes the discontinuity of $K_{x,-}$ and gives a substantial contribution to the equal point singularity of $K_{x,x}$ at short times. Nonetheless, even with these imperfections, experimental parameters can simply be obtained by fitting the theoretical predictions to the experimental curves.

\paragraph{Discussion --}

Forgetting about the equal point contributions, the correlation functions take the form of the trace of a succession of ``propagators'' $\Phi_{t_{k+1}-t_k}$ and superoperator insertions $c_k^{+}$ applied on the initial state. This form is extremely similar to that of expectation values of normal ordered products of local operators taken between continuous matrix product states (cMPS) \cite{verstraete2010,haegeman2013}, which are ansatz for states for 1-D quantum field theories. This is not a coincidence. There exists a mapping between bosonic cMPS and the state of the measurement field (or quantum noise) in continuous measurement setups \cite{haegeman2013,barrett2013}. Although we have favored a self-contained  stochastic proof and hinted at a discrete one, yet another derivation would have been possible using the language of cMPS and quantum noises.

This parallel with cMPS allows to learn how many correlation functions are typically needed to recover all system parameters. In the simple qubit example we considered, 2-point functions are clearly enough, but how far do we need to go in general? An analog of Wick's theorem by H\"ubener et al. \cite{hubener2013} shows that \emph{generically}, cMPS $N$-point functions can be reconstructed from 2 and 3 point functions. This means that, unless symmetries or numerical coincidences conspire, all the parameters of the system that are observable, i.e. in principle knowable from signal statistics, can be estimated using only 2 and 3 point functions.

Computing correlations functions with the exact formula is naturally still exponentially costly in the system size. However, provided the operators measured and the generators of the dynamics are quasi local with respect to a tensor decomposition into low dimensional factors, correlations functions can be efficiently approximated for short time differences $\Delta t$ using $\Phi_{\Delta t}\simeq \mathds{1} + \Delta t \mathscr{L} + \cdots$. This would suffice in most practical cases for parameter estimation. Importantly, for narrowly spaced time arguments, the contributions from the imperfect detection chain have a substantial impact on the statistics (see again Fig. \ref{fig:correlations}), and it is fortunate that they can be included via equation \eqref{eq:fullcorrelator}.

Finally, the exact expression \eqref{eq:fullcorrelator} possesses a rather simple structure allowing to extract properties of the signal that would not be obvious from the stochastic equations \eqref{eq:sme} and \eqref{eq:signal} we started with. For example one sees that correlation functions generically decrease exponentially fast with the time difference of their arguments. We can also readily see that correlators become independent from the initial state for large times if $\Phi$ has a unique stationary state. Hence in addition to its immediate interest for experimentalists, we can hope that the exact formula \eqref{eq:fullcorrelator} will be the basis of new mathematical developments in continuous measurement theory.

\begin{acknowledgments}
I am grateful to Alberto Barchielli, Denis Bernard, Michel Bauer, Geraldine Haack, Benjamin Huard, and Howard Wiseman for helpful suggestions. This work was made possible by support from the Alexander von Humboldt foundation and the Agence Nationale de la Recherche (ANR) contract ANR-14-CE25-0003-01.
\end{acknowledgments}

\bibliographystyle{apsrev4-1}
\bibliography{main}

\begin{thebibliography}{46}%
\makeatletter
\providecommand \@ifxundefined [1]{%
 \@ifx{#1\undefined}
}%
\providecommand \@ifnum [1]{%
 \ifnum #1\expandafter \@firstoftwo
 \else \expandafter \@secondoftwo
 \fi
}%
\providecommand \@ifx [1]{%
 \ifx #1\expandafter \@firstoftwo
 \else \expandafter \@secondoftwo
 \fi
}%
\providecommand \natexlab [1]{#1}%
\providecommand \enquote  [1]{``#1''}%
\providecommand \bibnamefont  [1]{#1}%
\providecommand \bibfnamefont [1]{#1}%
\providecommand \citenamefont [1]{#1}%
\providecommand \href@noop [0]{\@secondoftwo}%
\providecommand \href [0]{\begingroup \@sanitize@url \@href}%
\providecommand \@href[1]{\@@startlink{#1}\@@href}%
\providecommand \@@href[1]{\endgroup#1\@@endlink}%
\providecommand \@sanitize@url [0]{\catcode `\\12\catcode `\$12\catcode
  `\&12\catcode `\#12\catcode `\^12\catcode `\_12\catcode `\%12\relax}%
\providecommand \@@startlink[1]{}%
\providecommand \@@endlink[0]{}%
\providecommand \url  [0]{\begingroup\@sanitize@url \@url }%
\providecommand \@url [1]{\endgroup\@href {#1}{\urlprefix }}%
\providecommand \urlprefix  [0]{URL }%
\providecommand \Eprint [0]{\href }%
\providecommand \doibase [0]{http://dx.doi.org/}%
\providecommand \selectlanguage [0]{\@gobble}%
\providecommand \bibinfo  [0]{\@secondoftwo}%
\providecommand \bibfield  [0]{\@secondoftwo}%
\providecommand \translation [1]{[#1]}%
\providecommand \BibitemOpen [0]{}%
\providecommand \bibitemStop [0]{}%
\providecommand \bibitemNoStop [0]{.\EOS\space}%
\providecommand \EOS [0]{\spacefactor3000\relax}%
\providecommand \BibitemShut  [1]{\csname bibitem#1\endcsname}%
\let\auto@bib@innerbib\@empty
\bibitem [{\citenamefont {Attal}\ and\ \citenamefont
  {Pautrat}(2006)}]{attal2006}%
  \BibitemOpen
  \bibfield  {author} {\bibinfo {author} {\bibfnamefont {S.}~\bibnamefont
  {Attal}}\ and\ \bibinfo {author} {\bibfnamefont {Y.}~\bibnamefont
  {Pautrat}},\ }\href {\doibase 10.1007/s00023-005-0242-8} {\bibfield
  {journal} {\bibinfo  {journal} {Ann. H. Poincar\'e}\ }\textbf {\bibinfo
  {volume} {7}},\ \bibinfo {pages} {59} (\bibinfo {year} {2006})}\BibitemShut
  {NoStop}%
\bibitem [{\citenamefont {Attal}\ and\ \citenamefont
  {Pellegrini}(2010)}]{attal2010}%
  \BibitemOpen
  \bibfield  {author} {\bibinfo {author} {\bibfnamefont {S.}~\bibnamefont
  {Attal}}\ and\ \bibinfo {author} {\bibfnamefont {C.}~\bibnamefont
  {Pellegrini}},\ }\href {\doibase 10.1142/S1230161210000242} {\bibfield
  {journal} {\bibinfo  {journal} {Open Syst. Inf. Dyn.}\ }\textbf {\bibinfo
  {volume} {17}},\ \bibinfo {pages} {389} (\bibinfo {year} {2010})}\BibitemShut
  {NoStop}%
\bibitem [{\citenamefont {Mensky}(1979)}]{mensky1979}%
  \BibitemOpen
  \bibfield  {author} {\bibinfo {author} {\bibfnamefont {M.~B.}\ \bibnamefont
  {Mensky}},\ }\href {\doibase 10.1103/PhysRevD.20.384} {\bibfield  {journal}
  {\bibinfo  {journal} {Phys. Rev. D}\ }\textbf {\bibinfo {volume} {20}},\
  \bibinfo {pages} {384} (\bibinfo {year} {1979})}\BibitemShut {NoStop}%
\bibitem [{\citenamefont {Barchielli}\ \emph {et~al.}(1982)\citenamefont
  {Barchielli}, \citenamefont {Lanz},\ and\ \citenamefont
  {Prosperi}}]{barchielli1982}%
  \BibitemOpen
  \bibfield  {author} {\bibinfo {author} {\bibfnamefont {A.}~\bibnamefont
  {Barchielli}}, \bibinfo {author} {\bibfnamefont {L.}~\bibnamefont {Lanz}}, \
  and\ \bibinfo {author} {\bibfnamefont {G.~M.}\ \bibnamefont {Prosperi}},\
  }\href {\doibase 10.1007/BF02894935} {\bibfield  {journal} {\bibinfo
  {journal} {Il Nuovo Cimento B (1971-1996)}\ }\textbf {\bibinfo {volume}
  {72}},\ \bibinfo {pages} {79} (\bibinfo {year} {1982})}\BibitemShut {NoStop}%
\bibitem [{\citenamefont {Barchielli}(1986)}]{barchielli1986}%
  \BibitemOpen
  \bibfield  {author} {\bibinfo {author} {\bibfnamefont {A.}~\bibnamefont
  {Barchielli}},\ }\href {\doibase 10.1103/PhysRevA.34.1642} {\bibfield
  {journal} {\bibinfo  {journal} {Phys. Rev. A}\ }\textbf {\bibinfo {volume}
  {34}},\ \bibinfo {pages} {1642} (\bibinfo {year} {1986})}\BibitemShut
  {NoStop}%
\bibitem [{\citenamefont {Caves}(1986)}]{caves1986}%
  \BibitemOpen
  \bibfield  {author} {\bibinfo {author} {\bibfnamefont {C.~M.}\ \bibnamefont
  {Caves}},\ }\href {\doibase 10.1103/PhysRevD.33.1643} {\bibfield  {journal}
  {\bibinfo  {journal} {Phys. Rev. D}\ }\textbf {\bibinfo {volume} {33}},\
  \bibinfo {pages} {1643} (\bibinfo {year} {1986})}\BibitemShut {NoStop}%
\bibitem [{\citenamefont {Di\'osi}(1988)}]{diosi1988}%
  \BibitemOpen
  \bibfield  {author} {\bibinfo {author} {\bibfnamefont {L.}~\bibnamefont
  {Di\'osi}},\ }\href {\doibase https://doi.org/10.1016/0375-9601(88)90309-X}
  {\bibfield  {journal} {\bibinfo  {journal} {Phys. Lett. A}\ }\textbf
  {\bibinfo {volume} {129}},\ \bibinfo {pages} {419 } (\bibinfo {year}
  {1988})}\BibitemShut {NoStop}%
\bibitem [{\citenamefont {Barchielli}\ and\ \citenamefont
  {Belavkin}(1991)}]{barchielli1991}%
  \BibitemOpen
  \bibfield  {author} {\bibinfo {author} {\bibfnamefont {A.}~\bibnamefont
  {Barchielli}}\ and\ \bibinfo {author} {\bibfnamefont {V.~P.}\ \bibnamefont
  {Belavkin}},\ }\href {\doibase 10.1088/0305-4470/24/7/022} {\bibfield
  {journal} {\bibinfo  {journal} {J. Phys. A: Math. Gen.}\ }\textbf {\bibinfo
  {volume} {24}},\ \bibinfo {pages} {1495} (\bibinfo {year}
  {1991})}\BibitemShut {NoStop}%
\bibitem [{\citenamefont {Wiseman}\ and\ \citenamefont
  {Milburn}(1993)}]{wiseman1993}%
  \BibitemOpen
  \bibfield  {author} {\bibinfo {author} {\bibfnamefont {H.~M.}\ \bibnamefont
  {Wiseman}}\ and\ \bibinfo {author} {\bibfnamefont {G.~J.}\ \bibnamefont
  {Milburn}},\ }\href {\doibase 10.1103/PhysRevA.47.642} {\bibfield  {journal}
  {\bibinfo  {journal} {Phys. Rev. A}\ }\textbf {\bibinfo {volume} {47}},\
  \bibinfo {pages} {642} (\bibinfo {year} {1993})}\BibitemShut {NoStop}%
\bibitem [{\citenamefont {Wiseman}\ and\ \citenamefont
  {Milburn}(2009)}]{wiseman2009}%
  \BibitemOpen
  \bibfield  {author} {\bibinfo {author} {\bibfnamefont {H.~M.}\ \bibnamefont
  {Wiseman}}\ and\ \bibinfo {author} {\bibfnamefont {G.~J.}\ \bibnamefont
  {Milburn}},\ }\href@noop {} {\emph {\bibinfo {title} {Quantum measurement and
  control}}}\ (\bibinfo  {publisher} {Cambridge university press, Cambridge
  UK},\ \bibinfo {year} {2009})\BibitemShut {NoStop}%
\bibitem [{\citenamefont {Vijay}\ \emph {et~al.}(2012)\citenamefont {Vijay},
  \citenamefont {Macklin}, \citenamefont {Slichter}, \citenamefont {Weber},
  \citenamefont {Murch}, \citenamefont {Naik},\ and\ \citenamefont
  {Korotkov}}]{vijay2012}%
  \BibitemOpen
  \bibfield  {author} {\bibinfo {author} {\bibfnamefont {R.}~\bibnamefont
  {Vijay}}, \bibinfo {author} {\bibfnamefont {C.}~\bibnamefont {Macklin}},
  \bibinfo {author} {\bibfnamefont {D.~H.}\ \bibnamefont {Slichter}}, \bibinfo
  {author} {\bibfnamefont {S.~J.}\ \bibnamefont {Weber}}, \bibinfo {author}
  {\bibfnamefont {K.~W.}\ \bibnamefont {Murch}}, \bibinfo {author}
  {\bibfnamefont {R.}~\bibnamefont {Naik}}, \ and\ \bibinfo {author}
  {\bibfnamefont {I.}~\bibnamefont {Korotkov}, \bibfnamefont
  {A.~N.and~Siddiqi}},\ }\href {\doibase 10.1038/nature11505} {\bibfield
  {journal} {\bibinfo  {journal} {Nature}\ ,\ \bibinfo {pages} {77}} (\bibinfo
  {year} {2012})}\BibitemShut {NoStop}%
\bibitem [{\citenamefont {Katz}\ \emph {et~al.}(2006)\citenamefont {Katz},
  \citenamefont {Ansmann}, \citenamefont {Bialczak}, \citenamefont {Lucero},
  \citenamefont {McDermott}, \citenamefont {Neeley}, \citenamefont {Steffen},
  \citenamefont {Weig}, \citenamefont {Cleland}, \citenamefont {Martinis},\
  and\ \citenamefont {Korotkov}}]{katz2006}%
  \BibitemOpen
  \bibfield  {author} {\bibinfo {author} {\bibfnamefont {N.}~\bibnamefont
  {Katz}}, \bibinfo {author} {\bibfnamefont {M.}~\bibnamefont {Ansmann}},
  \bibinfo {author} {\bibfnamefont {R.~C.}\ \bibnamefont {Bialczak}}, \bibinfo
  {author} {\bibfnamefont {E.}~\bibnamefont {Lucero}}, \bibinfo {author}
  {\bibfnamefont {R.}~\bibnamefont {McDermott}}, \bibinfo {author}
  {\bibfnamefont {M.}~\bibnamefont {Neeley}}, \bibinfo {author} {\bibfnamefont
  {M.}~\bibnamefont {Steffen}}, \bibinfo {author} {\bibfnamefont {E.~M.}\
  \bibnamefont {Weig}}, \bibinfo {author} {\bibfnamefont {A.~N.}\ \bibnamefont
  {Cleland}}, \bibinfo {author} {\bibfnamefont {J.~M.}\ \bibnamefont
  {Martinis}}, \ and\ \bibinfo {author} {\bibfnamefont {A.~N.}\ \bibnamefont
  {Korotkov}},\ }\href {\doibase 10.1126/science.1126475} {\bibfield  {journal}
  {\bibinfo  {journal} {Science}\ }\textbf {\bibinfo {volume} {312}},\ \bibinfo
  {pages} {1498} (\bibinfo {year} {2006})}\BibitemShut {NoStop}%
\bibitem [{\citenamefont {Campagne-Ibarcq}\ \emph {et~al.}(2016)\citenamefont
  {Campagne-Ibarcq}, \citenamefont {Six}, \citenamefont {Bretheau},
  \citenamefont {Sarlette}, \citenamefont {Mirrahimi}, \citenamefont
  {Rouchon},\ and\ \citenamefont {Huard}}]{campagne2016}%
  \BibitemOpen
  \bibfield  {author} {\bibinfo {author} {\bibfnamefont {P.}~\bibnamefont
  {Campagne-Ibarcq}}, \bibinfo {author} {\bibfnamefont {P.}~\bibnamefont
  {Six}}, \bibinfo {author} {\bibfnamefont {L.}~\bibnamefont {Bretheau}},
  \bibinfo {author} {\bibfnamefont {A.}~\bibnamefont {Sarlette}}, \bibinfo
  {author} {\bibfnamefont {M.}~\bibnamefont {Mirrahimi}}, \bibinfo {author}
  {\bibfnamefont {P.}~\bibnamefont {Rouchon}}, \ and\ \bibinfo {author}
  {\bibfnamefont {B.}~\bibnamefont {Huard}},\ }\href {\doibase
  10.1103/PhysRevX.6.011002} {\bibfield  {journal} {\bibinfo  {journal} {Phys.
  Rev. X}\ }\textbf {\bibinfo {volume} {6}},\ \bibinfo {pages} {011002}
  (\bibinfo {year} {2016})}\BibitemShut {NoStop}%
\bibitem [{\citenamefont {R.}\ \emph {et~al.}(2005)\citenamefont {R.},
  \citenamefont {Stockton},\ and\ \citenamefont {Mabuchi}}]{vanhandel2005}%
  \BibitemOpen
  \bibfield  {author} {\bibinfo {author} {\bibfnamefont {V.~H.}\ \bibnamefont
  {R.}}, \bibinfo {author} {\bibfnamefont {J.~K.}\ \bibnamefont {Stockton}}, \
  and\ \bibinfo {author} {\bibfnamefont {H.}~\bibnamefont {Mabuchi}},\ }\href
  {\doibase 10.1088/1464-4266/7/10/001} {\bibfield  {journal} {\bibinfo
  {journal} {J. Opt. B Quantum Semiclassical Opt.}\ }\textbf {\bibinfo {volume}
  {7}},\ \bibinfo {pages} {S179} (\bibinfo {year} {2005})}\BibitemShut
  {NoStop}%
\bibitem [{\citenamefont {Jacobs}(2003)}]{jacobs2003}%
  \BibitemOpen
  \bibfield  {author} {\bibinfo {author} {\bibfnamefont {K.}~\bibnamefont
  {Jacobs}},\ }\href {\doibase 10.1103/PhysRevA.67.030301} {\bibfield
  {journal} {\bibinfo  {journal} {Phys. Rev. A}\ }\textbf {\bibinfo {volume}
  {67}},\ \bibinfo {pages} {030301} (\bibinfo {year} {2003})}\BibitemShut
  {NoStop}%
\bibitem [{\citenamefont {Combes}\ and\ \citenamefont
  {Jacobs}(2006)}]{combes2006}%
  \BibitemOpen
  \bibfield  {author} {\bibinfo {author} {\bibfnamefont {J.}~\bibnamefont
  {Combes}}\ and\ \bibinfo {author} {\bibfnamefont {K.}~\bibnamefont
  {Jacobs}},\ }\href {\doibase 10.1103/PhysRevLett.96.010504} {\bibfield
  {journal} {\bibinfo  {journal} {Phys. Rev. Lett.}\ }\textbf {\bibinfo
  {volume} {96}},\ \bibinfo {pages} {010504} (\bibinfo {year}
  {2006})}\BibitemShut {NoStop}%
\bibitem [{\citenamefont {Combes}\ \emph {et~al.}(2008)\citenamefont {Combes},
  \citenamefont {Wiseman},\ and\ \citenamefont {Jacobs}}]{combes2008}%
  \BibitemOpen
  \bibfield  {author} {\bibinfo {author} {\bibfnamefont {J.}~\bibnamefont
  {Combes}}, \bibinfo {author} {\bibfnamefont {H.~M.}\ \bibnamefont {Wiseman}},
  \ and\ \bibinfo {author} {\bibfnamefont {K.}~\bibnamefont {Jacobs}},\ }\href
  {\doibase 10.1103/PhysRevLett.100.160503} {\bibfield  {journal} {\bibinfo
  {journal} {Phys. Rev. Lett.}\ }\textbf {\bibinfo {volume} {100}},\ \bibinfo
  {pages} {160503} (\bibinfo {year} {2008})}\BibitemShut {NoStop}%
\bibitem [{\citenamefont {Combes}\ and\ \citenamefont
  {Wiseman}(2011)}]{combes2011}%
  \BibitemOpen
  \bibfield  {author} {\bibinfo {author} {\bibfnamefont {J.}~\bibnamefont
  {Combes}}\ and\ \bibinfo {author} {\bibfnamefont {H.~M.}\ \bibnamefont
  {Wiseman}},\ }\href {\doibase 10.1103/PhysRevX.1.011012} {\bibfield
  {journal} {\bibinfo  {journal} {Phys. Rev. X}\ }\textbf {\bibinfo {volume}
  {1}},\ \bibinfo {pages} {011012} (\bibinfo {year} {2011})}\BibitemShut
  {NoStop}%
\bibitem [{\citenamefont {Tilloy}(2016)}]{tilloy2016}%
  \BibitemOpen
  \bibfield  {author} {\bibinfo {author} {\bibfnamefont {A.}~\bibnamefont
  {Tilloy}},\ }\href {\doibase 10.1103/PhysRevA.93.052309} {\bibfield
  {journal} {\bibinfo  {journal} {Phys. Rev. A}\ }\textbf {\bibinfo {volume}
  {93}},\ \bibinfo {pages} {052309} (\bibinfo {year} {2016})}\BibitemShut
  {NoStop}%
\bibitem [{\citenamefont {Ahn}\ \emph {et~al.}(2002)\citenamefont {Ahn},
  \citenamefont {Doherty},\ and\ \citenamefont {Landahl}}]{ahn2002}%
  \BibitemOpen
  \bibfield  {author} {\bibinfo {author} {\bibfnamefont {C.}~\bibnamefont
  {Ahn}}, \bibinfo {author} {\bibfnamefont {A.~C.}\ \bibnamefont {Doherty}}, \
  and\ \bibinfo {author} {\bibfnamefont {A.~J.}\ \bibnamefont {Landahl}},\
  }\href {\doibase 10.1103/PhysRevA.65.042301} {\bibfield  {journal} {\bibinfo
  {journal} {Phys. Rev. A}\ }\textbf {\bibinfo {volume} {65}},\ \bibinfo
  {pages} {042301} (\bibinfo {year} {2002})}\BibitemShut {NoStop}%
\bibitem [{\citenamefont {Atalaya}\ \emph
  {et~al.}(2017{\natexlab{a}})\citenamefont {Atalaya}, \citenamefont {Bahrami},
  \citenamefont {Pryadko},\ and\ \citenamefont {Korotkov}}]{atalaya2017_0}%
  \BibitemOpen
  \bibfield  {author} {\bibinfo {author} {\bibfnamefont {J.}~\bibnamefont
  {Atalaya}}, \bibinfo {author} {\bibfnamefont {M.}~\bibnamefont {Bahrami}},
  \bibinfo {author} {\bibfnamefont {L.~P.}\ \bibnamefont {Pryadko}}, \ and\
  \bibinfo {author} {\bibfnamefont {A.~N.}\ \bibnamefont {Korotkov}},\ }\href
  {\doibase 10.1103/PhysRevA.95.032317} {\bibfield  {journal} {\bibinfo
  {journal} {Phys. Rev. A}\ }\textbf {\bibinfo {volume} {95}},\ \bibinfo
  {pages} {032317} (\bibinfo {year} {2017}{\natexlab{a}})}\BibitemShut
  {NoStop}%
\bibitem [{\citenamefont {Chase}\ and\ \citenamefont
  {Geremia}(2009)}]{chase2009}%
  \BibitemOpen
  \bibfield  {author} {\bibinfo {author} {\bibfnamefont {B.~A.}\ \bibnamefont
  {Chase}}\ and\ \bibinfo {author} {\bibfnamefont {J.~M.}\ \bibnamefont
  {Geremia}},\ }\href {\doibase 10.1103/PhysRevA.79.022314} {\bibfield
  {journal} {\bibinfo  {journal} {Phys. Rev. A}\ }\textbf {\bibinfo {volume}
  {79}},\ \bibinfo {pages} {022314} (\bibinfo {year} {2009})}\BibitemShut
  {NoStop}%
\bibitem [{\citenamefont {Gambetta}\ and\ \citenamefont
  {Wiseman}(2001)}]{gambetta2001}%
  \BibitemOpen
  \bibfield  {author} {\bibinfo {author} {\bibfnamefont {J.}~\bibnamefont
  {Gambetta}}\ and\ \bibinfo {author} {\bibfnamefont {H.~M.}\ \bibnamefont
  {Wiseman}},\ }\href {\doibase 10.1103/PhysRevA.64.042105} {\bibfield
  {journal} {\bibinfo  {journal} {Phys. Rev. A}\ }\textbf {\bibinfo {volume}
  {64}},\ \bibinfo {pages} {042105} (\bibinfo {year} {2001})}\BibitemShut
  {NoStop}%
\bibitem [{\citenamefont {Six}\ \emph {et~al.}(2015)\citenamefont {Six},
  \citenamefont {Campagne-Ibarcq}, \citenamefont {Bretheau}, \citenamefont
  {Huard},\ and\ \citenamefont {Rouchon}}]{six2015}%
  \BibitemOpen
  \bibfield  {author} {\bibinfo {author} {\bibfnamefont {P.}~\bibnamefont
  {Six}}, \bibinfo {author} {\bibfnamefont {P.}~\bibnamefont
  {Campagne-Ibarcq}}, \bibinfo {author} {\bibfnamefont {L.}~\bibnamefont
  {Bretheau}}, \bibinfo {author} {\bibfnamefont {B.}~\bibnamefont {Huard}}, \
  and\ \bibinfo {author} {\bibfnamefont {P.}~\bibnamefont {Rouchon}},\ }in\
  \href {\doibase 10.1109/CDC.2015.7403443} {\emph {\bibinfo {booktitle} {54th
  IEEE Conference on Decision and Control (CDC)}}}\ (\bibinfo {year} {2015})\
  pp.\ \bibinfo {pages} {7742--7748}\BibitemShut {NoStop}%
\bibitem [{\citenamefont {Jordan}\ and\ \citenamefont
  {B\"uttiker}(2005)}]{jordan2005}%
  \BibitemOpen
  \bibfield  {author} {\bibinfo {author} {\bibfnamefont {A.~N.}\ \bibnamefont
  {Jordan}}\ and\ \bibinfo {author} {\bibfnamefont {M.}~\bibnamefont
  {B\"uttiker}},\ }\href {\doibase 10.1103/PhysRevLett.95.220401} {\bibfield
  {journal} {\bibinfo  {journal} {Phys. Rev. Lett.}\ }\textbf {\bibinfo
  {volume} {95}},\ \bibinfo {pages} {220401} (\bibinfo {year}
  {2005})}\BibitemShut {NoStop}%
\bibitem [{\citenamefont {Atalaya}\ \emph
  {et~al.}(2017{\natexlab{b}})\citenamefont {Atalaya}, \citenamefont
  {Hacohen-Gourgy}, \citenamefont {Martin}, \citenamefont {Siddiqi},\ and\
  \citenamefont {Korotkov}}]{atalaya2017_1}%
  \BibitemOpen
  \bibfield  {author} {\bibinfo {author} {\bibfnamefont {J.}~\bibnamefont
  {Atalaya}}, \bibinfo {author} {\bibfnamefont {S.}~\bibnamefont
  {Hacohen-Gourgy}}, \bibinfo {author} {\bibfnamefont {L.~S.}\ \bibnamefont
  {Martin}}, \bibinfo {author} {\bibfnamefont {I.}~\bibnamefont {Siddiqi}}, \
  and\ \bibinfo {author} {\bibfnamefont {A.~N.}\ \bibnamefont {Korotkov}},\
  }\href {https://arxiv.org/abs/1702.08077} {\bibfield  {journal} {\bibinfo
  {journal} {arXiv:1702.08077}\ } (\bibinfo {year}
  {2017}{\natexlab{b}})}\BibitemShut {NoStop}%
\bibitem [{\citenamefont {Zoller}\ and\ \citenamefont
  {Gardiner}(1997)}]{zoller1997}%
  \BibitemOpen
  \bibfield  {author} {\bibinfo {author} {\bibfnamefont {P.}~\bibnamefont
  {Zoller}}\ and\ \bibinfo {author} {\bibfnamefont {C.~W.}\ \bibnamefont
  {Gardiner}},\ }in\ \href@noop {} {\emph {\bibinfo {booktitle} {Lecture Notes
  for the Les Houches Summer School LXIII on Quantum Fluctuations in July 1995,
  Edited by E. Giacobino and S. Reynaud}}}\ (\bibinfo  {publisher} {Elsevier
  Science Publishers B.V.},\ \bibinfo {year} {1997})\BibitemShut {NoStop}%
\bibitem [{\citenamefont {Barchielli}(2006)}]{barchielli2006}%
  \BibitemOpen
  \bibfield  {author} {\bibinfo {author} {\bibfnamefont {A.}~\bibnamefont
  {Barchielli}},\ }in\ \href {http://www.springer.com/gp/book/9783540309932}
  {\emph {\bibinfo {booktitle} {Open Quantum Systems III}}},\ \bibinfo {editor}
  {edited by\ \bibinfo {editor} {\bibfnamefont {S.}~\bibnamefont {Attal}},
  \bibinfo {editor} {\bibfnamefont {A.}~\bibnamefont {Joye}}, \ and\ \bibinfo
  {editor} {\bibfnamefont {C.-A.}\ \bibnamefont {Pillet}}}\ (\bibinfo
  {publisher} {Springer},\ \bibinfo {address} {Berlin},\ \bibinfo {year}
  {2006})\ pp.\ \bibinfo {pages} {207--291}\BibitemShut {NoStop}%
\bibitem [{\citenamefont {Barchielli}\ and\ \citenamefont
  {Gregoratti}(2009)}]{barchielli2009}%
  \BibitemOpen
  \bibfield  {author} {\bibinfo {author} {\bibfnamefont {A.}~\bibnamefont
  {Barchielli}}\ and\ \bibinfo {author} {\bibfnamefont {M.}~\bibnamefont
  {Gregoratti}},\ }\href {http://www.springer.com/us/book/9783642012976} {\emph
  {\bibinfo {title} {Quantum trajectories and measurements in continuous time:
  the diffusive case}}},\ Vol.\ \bibinfo {volume} {782}\ (\bibinfo  {publisher}
  {Springer-Verlag Berlin Heidelberg},\ \bibinfo {year} {2009})\BibitemShut
  {NoStop}%
\bibitem [{\citenamefont {Di\'osi}(2016)}]{diosi2016}%
  \BibitemOpen
  \bibfield  {author} {\bibinfo {author} {\bibfnamefont {L.}~\bibnamefont
  {Di\'osi}},\ }\href {\doibase 10.1103/PhysRevA.94.010103} {\bibfield
  {journal} {\bibinfo  {journal} {Phys. Rev. A}\ }\textbf {\bibinfo {volume}
  {94}},\ \bibinfo {pages} {010103} (\bibinfo {year} {2016})}\BibitemShut
  {NoStop}%
\bibitem [{\citenamefont {Atalaya}\ \emph {et~al.}(2018)\citenamefont
  {Atalaya}, \citenamefont {Hacohen-Gourgy}, \citenamefont {Martin},
  \citenamefont {Siddiqi},\ and\ \citenamefont {Korotkov}}]{atalaya2017_2}%
  \BibitemOpen
  \bibfield  {author} {\bibinfo {author} {\bibfnamefont {J.}~\bibnamefont
  {Atalaya}}, \bibinfo {author} {\bibfnamefont {S.}~\bibnamefont
  {Hacohen-Gourgy}}, \bibinfo {author} {\bibfnamefont {L.~S.}\ \bibnamefont
  {Martin}}, \bibinfo {author} {\bibfnamefont {I.}~\bibnamefont {Siddiqi}}, \
  and\ \bibinfo {author} {\bibfnamefont {A.~N.}\ \bibnamefont {Korotkov}},\
  }\href {\doibase 10.1103/PhysRevA.97.020104} {\bibfield  {journal} {\bibinfo
  {journal} {Phys. Rev. A}\ }\textbf {\bibinfo {volume} {97}},\ \bibinfo
  {pages} {020104} (\bibinfo {year} {2018})}\BibitemShut {NoStop}%
\bibitem [{Note1()}]{Note1}%
  \BibitemOpen
  \bibinfo {note} {The SME can actually take an even more general form (see
  \protect \textit {e.g.} \cite {wiseman2001,barchielli1993,barchielli1995}).
  Our derivation is straightforward to extend to such settings.}\BibitemShut
  {Stop}%
\bibitem [{\citenamefont {Jacobs}\ and\ \citenamefont
  {Steck}(2006)}]{jacobs2006}%
  \BibitemOpen
  \bibfield  {author} {\bibinfo {author} {\bibfnamefont {K.}~\bibnamefont
  {Jacobs}}\ and\ \bibinfo {author} {\bibfnamefont {D.~A.}\ \bibnamefont
  {Steck}},\ }\href {\doibase 10.1080/00107510601101934} {\bibfield  {journal}
  {\bibinfo  {journal} {Contemp. Phys.}\ }\textbf {\bibinfo {volume} {47}},\
  \bibinfo {pages} {279} (\bibinfo {year} {2006})}\BibitemShut {NoStop}%
\bibitem [{Note2()}]{Note2}%
  \BibitemOpen
  \bibinfo {note} {We have taken the convention of \cite {jacobs2006} which
  means there is a factor of $2$ difference with the definition of the signal
  in \cite {atalaya2017_2}.}\BibitemShut {Stop}%
\bibitem [{\citenamefont {{\O}ksendal}(2003)}]{oksendal2003}%
  \BibitemOpen
  \bibfield  {author} {\bibinfo {author} {\bibfnamefont {B.}~\bibnamefont
  {{\O}ksendal}},\ }\href@noop {} {\emph {\bibinfo {title} {Stochastic
  differential equations}}}\ (\bibinfo  {publisher} {Springer, Berlin
  Heidelberg},\ \bibinfo {year} {2003})\BibitemShut {NoStop}%
\bibitem [{Note3()}]{Note3}%
  \BibitemOpen
  \bibinfo {note} {This is sometimes done the other way around (see e.g. \cite
  {jacobs2006,barchielli2009}). Starting from a linear SME with Gaussian
  signal, one derives the non-linear SME \protect \textup {\hbox {\mathsurround
  \z@ \protect \normalfont (\ignorespaces \ref {eq:sme}\unskip \@@italiccorr
  )}} and drifted signal \protect \textup {\hbox {\mathsurround \z@ \protect
  \normalfont (\ignorespaces \ref {eq:signal}\unskip \@@italiccorr )}}
  normalizing the solution and changing the probability measure.}\BibitemShut
  {Stop}%
\bibitem [{\citenamefont {Bauer}\ \emph {et~al.}(2014)\citenamefont {Bauer},
  \citenamefont {Bernard},\ and\ \citenamefont {Tilloy}}]{tilloy2014}%
  \BibitemOpen
  \bibfield  {author} {\bibinfo {author} {\bibfnamefont {M.}~\bibnamefont
  {Bauer}}, \bibinfo {author} {\bibfnamefont {D.}~\bibnamefont {Bernard}}, \
  and\ \bibinfo {author} {\bibfnamefont {A.}~\bibnamefont {Tilloy}},\ }\href
  {\doibase 10.1088/1742-5468/2014/09/P09001} {\bibfield  {journal} {\bibinfo
  {journal} {J. Stat. Mech.}\ }\textbf {\bibinfo {volume} {2014}},\ \bibinfo
  {pages} {P09001} (\bibinfo {year} {2014})}\BibitemShut {NoStop}%
\bibitem [{\citenamefont {Gross}\ \emph {et~al.}(2018)\citenamefont {Gross},
  \citenamefont {Caves}, \citenamefont {Milburn},\ and\ \citenamefont
  {Combes}}]{gross2018}%
  \BibitemOpen
  \bibfield  {author} {\bibinfo {author} {\bibfnamefont {J.~A.}\ \bibnamefont
  {Gross}}, \bibinfo {author} {\bibfnamefont {C.~M.}\ \bibnamefont {Caves}},
  \bibinfo {author} {\bibfnamefont {G.~J.}\ \bibnamefont {Milburn}}, \ and\
  \bibinfo {author} {\bibfnamefont {J.}~\bibnamefont {Combes}},\ }\href
  {\doibase 10.1088/2058-9565/aaa39f} {\bibfield  {journal} {\bibinfo
  {journal} {Quantum Sci. Technol.}\ }\textbf {\bibinfo {volume} {3}},\
  \bibinfo {pages} {024005} (\bibinfo {year} {2018})}\BibitemShut {NoStop}%
\bibitem [{\citenamefont {Ficheux}\ \emph {et~al.}(2017)\citenamefont
  {Ficheux}, \citenamefont {Jezouin}, \citenamefont {Leghtas},\ and\
  \citenamefont {Huard}}]{ficheux2017}%
  \BibitemOpen
  \bibfield  {author} {\bibinfo {author} {\bibfnamefont {Q.}~\bibnamefont
  {Ficheux}}, \bibinfo {author} {\bibfnamefont {S.}~\bibnamefont {Jezouin}},
  \bibinfo {author} {\bibfnamefont {Z.}~\bibnamefont {Leghtas}}, \ and\
  \bibinfo {author} {\bibfnamefont {B.}~\bibnamefont {Huard}},\ }\href
  {https://arxiv.org/abs/1711.01208} {\bibfield  {journal} {\bibinfo  {journal}
  {arXiv:1711.01208}\ } (\bibinfo {year} {2017})}\BibitemShut {NoStop}%
\bibitem [{\citenamefont {Verstraete}\ and\ \citenamefont
  {Cirac}(2010)}]{verstraete2010}%
  \BibitemOpen
  \bibfield  {author} {\bibinfo {author} {\bibfnamefont {F.}~\bibnamefont
  {Verstraete}}\ and\ \bibinfo {author} {\bibfnamefont {J.~I.}\ \bibnamefont
  {Cirac}},\ }\href {\doibase 10.1103/PhysRevLett.104.190405} {\bibfield
  {journal} {\bibinfo  {journal} {Phys. Rev. Lett.}\ }\textbf {\bibinfo
  {volume} {104}},\ \bibinfo {pages} {190405} (\bibinfo {year}
  {2010})}\BibitemShut {NoStop}%
\bibitem [{\citenamefont {Haegeman}\ \emph {et~al.}(2013)\citenamefont
  {Haegeman}, \citenamefont {Cirac}, \citenamefont {Osborne},\ and\
  \citenamefont {Verstraete}}]{haegeman2013}%
  \BibitemOpen
  \bibfield  {author} {\bibinfo {author} {\bibfnamefont {J.}~\bibnamefont
  {Haegeman}}, \bibinfo {author} {\bibfnamefont {J.~I.}\ \bibnamefont {Cirac}},
  \bibinfo {author} {\bibfnamefont {T.~J.}\ \bibnamefont {Osborne}}, \ and\
  \bibinfo {author} {\bibfnamefont {F.}~\bibnamefont {Verstraete}},\ }\href
  {\doibase 10.1103/PhysRevB.88.085118} {\bibfield  {journal} {\bibinfo
  {journal} {Phys. Rev. B}\ }\textbf {\bibinfo {volume} {88}},\ \bibinfo
  {pages} {085118} (\bibinfo {year} {2013})}\BibitemShut {NoStop}%
\bibitem [{\citenamefont {Barrett}\ \emph {et~al.}(2013)\citenamefont
  {Barrett}, \citenamefont {Hammerer}, \citenamefont {Harrison}, \citenamefont
  {Northup},\ and\ \citenamefont {Osborne}}]{barrett2013}%
  \BibitemOpen
  \bibfield  {author} {\bibinfo {author} {\bibfnamefont {S.}~\bibnamefont
  {Barrett}}, \bibinfo {author} {\bibfnamefont {K.}~\bibnamefont {Hammerer}},
  \bibinfo {author} {\bibfnamefont {S.}~\bibnamefont {Harrison}}, \bibinfo
  {author} {\bibfnamefont {T.~E.}\ \bibnamefont {Northup}}, \ and\ \bibinfo
  {author} {\bibfnamefont {T.~J.}\ \bibnamefont {Osborne}},\ }\href {\doibase
  10.1103/PhysRevLett.110.090501} {\bibfield  {journal} {\bibinfo  {journal}
  {Phys. Rev. Lett.}\ }\textbf {\bibinfo {volume} {110}},\ \bibinfo {pages}
  {090501} (\bibinfo {year} {2013})}\BibitemShut {NoStop}%
\bibitem [{\citenamefont {H\"ubener}\ \emph {et~al.}(2013)\citenamefont
  {H\"ubener}, \citenamefont {Mari},\ and\ \citenamefont
  {Eisert}}]{hubener2013}%
  \BibitemOpen
  \bibfield  {author} {\bibinfo {author} {\bibfnamefont {R.}~\bibnamefont
  {H\"ubener}}, \bibinfo {author} {\bibfnamefont {A.}~\bibnamefont {Mari}}, \
  and\ \bibinfo {author} {\bibfnamefont {J.}~\bibnamefont {Eisert}},\ }\href
  {\doibase 10.1103/PhysRevLett.110.040401} {\bibfield  {journal} {\bibinfo
  {journal} {Phys. Rev. Lett.}\ }\textbf {\bibinfo {volume} {110}},\ \bibinfo
  {pages} {040401} (\bibinfo {year} {2013})}\BibitemShut {NoStop}%
\bibitem [{\citenamefont {Wiseman}\ and\ \citenamefont
  {Di\'osi}(2001)}]{wiseman2001}%
  \BibitemOpen
  \bibfield  {author} {\bibinfo {author} {\bibfnamefont {H.}~\bibnamefont
  {Wiseman}}\ and\ \bibinfo {author} {\bibfnamefont {L.}~\bibnamefont
  {Di\'osi}},\ }\href {\doibase https://doi.org/10.1016/S0301-0104(01)00296-8}
  {\bibfield  {journal} {\bibinfo  {journal} {Chem. Phys.}\ }\textbf {\bibinfo
  {volume} {268}},\ \bibinfo {pages} {91 } (\bibinfo {year}
  {2001})}\BibitemShut {NoStop}%
\bibitem [{\citenamefont {Barchielli}\ \emph {et~al.}(1993)\citenamefont
  {Barchielli}, \citenamefont {Holevo},\ and\ \citenamefont
  {Lupieri}}]{barchielli1993}%
  \BibitemOpen
  \bibfield  {author} {\bibinfo {author} {\bibfnamefont {A.}~\bibnamefont
  {Barchielli}}, \bibinfo {author} {\bibfnamefont {A.~S.}\ \bibnamefont
  {Holevo}}, \ and\ \bibinfo {author} {\bibfnamefont {G.}~\bibnamefont
  {Lupieri}},\ }\href {\doibase 10.1007/BF01047573} {\bibfield  {journal}
  {\bibinfo  {journal} {J. Theor. Probab.}\ }\textbf {\bibinfo {volume} {6}},\
  \bibinfo {pages} {231} (\bibinfo {year} {1993})}\BibitemShut {NoStop}%
\bibitem [{\citenamefont {Barchielli}\ and\ \citenamefont
  {Holevo}(1995)}]{barchielli1995}%
  \BibitemOpen
  \bibfield  {author} {\bibinfo {author} {\bibfnamefont {A.}~\bibnamefont
  {Barchielli}}\ and\ \bibinfo {author} {\bibfnamefont {A.}~\bibnamefont
  {Holevo}},\ }\href {\doibase https://doi.org/10.1016/0304-4149(95)00011-U}
  {\bibfield  {journal} {\bibinfo  {journal} {Stoch. Process. Their. Appl.}\
  }\textbf {\bibinfo {volume} {58}},\ \bibinfo {pages} {293 } (\bibinfo {year}
  {1995})}\BibitemShut {NoStop}%
\end{thebibliography}%

\end{document}